\documentclass[fleqn,twoside]{article}
\usepackage{espcrc2}

\newcommand{\AmS}{{\protect\the\textfont2
  A\kern-.1667em\lower.5ex\hbox{M}\kern-.125emS}}

\title{Prospects of the Zee model}

\author{Yoshio Koide  \address{ Department of Physics, 
University of Shizuoka, 52-1 Yada, Shizuoka, Japan 422-8526} 
        \thanks{E-mail address: koide@u-shizuoka-ken.ac.jp }
}

\begin{document}

\begin{abstract}
The Zee model is one of promising models of neutrino mass 
generation mechanism. However, 
 the original Zee model is not on the framework of 
the ground unification scenario, and moreover, 
it is recently pointed out that the 
predicted value of $\sin^2 2\theta_{solar}$ must be satisfied 
the relation $\sin^2 2\theta_{solar} > 0.99$. 
We discuss whether 
possible GUT versions of the Zee model can be free from the 
severe constraint $\sin^2 2\theta_{solar} >0.99$ or not. 
We will conclude that the following two models are
promising: an $R$-parity violating SUSY GUT model and
an SO(10) model with a 126-plet scalar.
\vspace{1pc}
\end{abstract}

\maketitle

\section{INTRODUCTION}

The Zee model \cite{Zee} is one of the attractive models 
for neutrino mass matrix. 
The mass matrix form is given by 
$$
M_\nu = \left(\begin{array}{ccc}
0 & a & c \\
a & 0 & b \\
c & b & 0
\end{array} \right)\  ,
\eqno(1.1)
$$
where 
$$
 a=f_{e \mu} (m_\mu^2 - m_e^2) K \ ,
$$
$$
 b=f_{\mu \tau}(m_\tau^2 - m_\mu^2) K \ , 
\eqno(1.2)
$$
$$
c=f_{\tau e} (m_e^2 - m_\tau^2) K \ ,
$$
and $K$ is a common factor. The model has only 3 free parameters and it 
can naturally lead to a large neutrino mixing \cite{Zee2}.
Especially, for the case 
$a=c\gg b$,
it leads to a { bi-maximal mixing} \cite{Jarlskog}
$$
U \simeq 
\left( \begin{array}{ccc}
\frac{1}{\sqrt{2}} & -\frac{1}{\sqrt{2}} & 0 \\
\frac{1}{2} & \frac{1}{2} & -\frac{1}{\sqrt{2}} \\
\frac{1}{2} & \frac{1}{2} & \frac{1}{\sqrt{2}}
\end{array} \right)\ ,
\eqno(1.3)
$$
with $\Delta m_{12}^2/\Delta m_{23}^2
\simeq \sqrt{2} {b}/{a}$. 

However, from the standpoint of the unification of quarks and leptons, 
there are some problems to be overcome.

(1) The masses of the quarks and charged leptons are generated by the
Higgs mechanism, so that they depend on the Yukawa 
coupling constants $y_{ij}$, while only neutrino masses are 
generated radiatively, so that they depend on the Zee coupling 
constants $f_{ij}$.
In general, $f_{ij}$ are independent of $y_{ij}$.
We must seek for a principle which relates $f_{ij}$ to $y_{ij}$.
(For such an attempt, for example, see Ref.\cite{Koide-Ghosal},)

(2) The Zee model is not embedded in a Grand Unification
Theory (GUT).   We must seek for a room for the Zee scalar $h^+$
in a GUT scenario.

(3) The Zee model leads to  a severe constraint \cite{Koide}
$\sin^2 2\theta_{solar} >0.99$ under the observed ratio
$\Delta m^2_{solar}/\Delta m_{atm}^2\ll 1$.
The present solar neutrino data \cite{solar} show
$\sin^2 2\theta_{solar} \sim 0.8$.
We must investigate a modified Zee model which can be free from
the severe constraint $\sin^2 2\theta_{solar} >0.99$.

The problem (3) has recently pointed out by the author\cite{Koide}. 
A parameter independent investigation leads to 
a severe constraint on the value of
$ \sin^2 2\theta_{solar}$
$$
\sin^2 2\theta_{solar} \geq
1 - \frac{1}{16}\left(
\frac{\Delta m^2_{solar}}{\Delta m_{atm}^2}\right)^2.
\eqno(1.4)
$$
The conclusion cannot be loosened 
even if we take RGE effects
into consideration:
Also, we can show that the two-loop effect 
is negligibly small.

The simple ways to escape from the constraint (1.8) will by as follows:
One is to consider \cite{ZeeH2} that the Yukawa vertices of 
the charged leptons can couple to both scalars
$\phi_1$ and $\phi_2$.
Another one \cite{Zee_nuR} is to introduce a single 
right-handed neutrino $\nu_R$ and
a second singlet Zee scalar $S^+$.
 Also, a model with a new doubly charged scalar $k^{++}$ is 
interesting because the two loop 
effects in such a model can give non-negligible contributions to the neutrino
masses \cite{k++}. 
As another attractive model, there is an idea \cite{D-P-T,R_SUSY}
that in an $R$-parity violating SUSY model
we identify Zee scalar  $h^+$ as slepton $ \tilde{e}_R$. Then, we can obtain
 additional contributions from
$d$-quark loops to the neutrino masses. 
However, if we want to extend the model to a GUT scenario, we will meet a new 
trouble ``proton decay" as stated in the next section.

Anyhow, these models are not connected to GUT scenarios.
We want to embed the Zee model into a GUT scenario.
For an extended Zee model based on a GUT scenario, there is,
for example, the Haba-Matsuda-Tanimoto
model \cite{Haba}. 
They have regarded the Zee scalar $h^+$ as 
a member of the
messenger field $M_{10} +\overline{M_{10}}$ of SUSY-breaking
on the basis of an SU(5) SUSY GUT.
However, their model cannot escape from the constraint (1.8)
as we discuss in the next section.

\section{EXTENDED ZEE MODEL WITH GUT}


We identify the Zee scalar $h^+$ as a member of SU(5) 10-plet
scalar (including a case of $\overline{5}+10$ sfermion).
Then, we must pay attention to the following items:

(a) Proton decay is safely forbidden.

(b) The model is free from the severe constraint 
$\sin^2 2\theta_{solar} > 0.99$ in the Zee model.

\subsection{How to avoid proton decay}

Generally, the $\overline{5}$-plet and 10-plet scalars 
$\phi_{\overline{5}}$ and $\phi_{10}$  can couple 
to the $\overline{5}+10$-plet fermions as follows.
$$
\begin{array}{l}
A_5 \equiv
(\overline\psi^c_{\overline{5}})^A (\psi_{10})_{AB} 
(\phi_{\overline{5}})^B \ , \\
A_{10} \equiv 
({\overline{\psi}}^c_{\overline{5}})^A (\psi_{\overline{5}})^B
 (\phi_{10})_{AB} \ , \\
B_5 \equiv \varepsilon^{ABCDE} ({\overline{\psi}}^c_{10})_{AB}
  (\psi_{10})_{CD} ({\overline{\phi}}_{\overline{5}})_E \ .\\
\end{array} 
\eqno(2.1)
$$
The terms $A_{10}$ can contribute to the radiative neutrino mass,
while the terms $A_5$ and $B_5$ induce the proton decay.
We want the terms $A_{10}$, but do not want  terms $A_{5}$ and $B_{5}$.

The conventional $R$-parity violating SUSY model \cite{R_SUSY} 
contains terms $A_{10}$, but 
also contains terms $A_{5}$ and $B_{5}$, so that the model is ruled out
from the candidates of the extended Zee model based on a GUT scenario.

For example, in the Haba-Matsuda-Tanimoto mode \cite{Haba}, 
there is no SU(5) 5-plet scalar 
(except for the conventional Higgs scalars), so that there is
no proton decay due to $A_5$ and $B_5$ terms.
In their model, in principle, the down-quark loop diagrams can 
contribute to the neutrino masses in addition to the charged
lepton loop diagrams.  Such the down-quark loop diagrams can
contribute to the non-diagonal elements of the neutrino mass
matrix.
However, the down-quark loop diagrams
include a colored Higgs scalar (the triplet component of the SU(5) 
5-plet Higgs scalar). We suppose that the colored Higgs scalar
has a mass of the order of the grand unification scale in order
to suppress the proton decay due to the colored Higgs scalar.
Therefore, the contributions are negligibly small, 
so that the model
cannot still be free from the severe constraint
$\sin^2 2\theta_{solar}>0.99$. 

\subsection{$R$-parity violating model}

Note that if we assume that $R$-parity violating
interactions are allowed only for special generations
(families), then we can choose such the $A_5$ 
and $B_5$ interactions as they do not contribute to
the proton decay without breaking the SU(5) GUT.
\vglue.1in

For example, we assume
$$
\begin{array}{l}
A_5 \equiv \lambda_{ij}^{3}
(\overline\psi^c_{\overline{5}})^A_{i} 
(\psi_{10})_{{3}AB} 
(\widetilde{\psi}_{\overline{5}})^B_{j} \ ,\\
A_{10} \equiv \lambda_{ij}^3
({\overline{\psi}}^c_{\overline{5}})^A_i 
(\psi_{\overline{5}})^B_{j}
 (\widetilde{\psi}_{10})_{{3}AB} \ ,
\end{array}
\eqno(2.4)
$$
i.e., only the interactions with the SU(5) 10-plet superfield 
of the third generation violate the $R$-parity.
Besides, we must assume this expression (5.4) is true on the basis
on which the up-quark mass matrix is diagonal, because if it is not so, 
the proton decay still occurs through the CKM mixing 
$V^*_{13}$.
In order to forbid the $u$-$t$ mixing at energy scale, we must
assume additional flavor symmetries (a discrete symmetry, and so on).


Although the assumption (2.4) is somewhat unnatural,
anyhow, the model can give nonvanishing diagonal
elements in the neutrino mass matrix.
The contributions give the form 
$$
M_{ij}^{e,d\ loop} = m_0 f_i f_j
\ \ \ \ {\rm with}\ f_i =\lambda^3_{i3}
\ .
\eqno(2.5)
$$
This matrix form (2.5) corresponds to the 
Drees-Pakvasa-Tata-Veldhuis model \cite{D-P-T}.

\subsection{Model with a SU(2) triplet scalar}

Except for the ``restricted" R-parity violating SUSY GUT model,
we have failed to build a GUT scenario in which 
the original Zee scalar $h^+$ is embedded into an
SU(5) 10 scalar (or SO(10) 120 scalar), as far as
we adhere to the model-building without the
constraint $\sin^2 2\theta_{solar}>0.99$.

Alternatively, we consider 
an SU(2) triplet scalar
$\phi=(\phi^{++}, \phi^{+}, \phi^0)$.
Usually, such a Higgs scalar can directly give
Majorana masses for the left-handed neutrinos $\nu_L$.
We assume that the potential for $\phi$ does not 
have a minimum.
Then, 
we obtain the following radiative masses
$$
M_{ij} = f_{ij} \left( (m^e_i)^2 + (m^e_j)^2
\right) K
\ ,
\eqno(2.6)
$$
instead of the Zee mass matrix (1.1) with (1.2).

We regard the triplet scalar $\phi$ as a member of 
the SO(10) 126-plet scalar
$$
\begin{array}{l}
126= 1_{-10}+\overline{5}_{-2}+ 10_{-6}+
{\overline{15}_{+6}}
+45_{+2} +\overline{50}_{-2} \ ,\\
\ \ \ \ \ \ \ \ \ \ \ \ \ \ \ \ \ \ \
\end{array}
\eqno(2.7)
$$
of $SU(5)\times U(1)$,where the triplet scalar $\phi$ belongs to the 
component $\overline{15}_{+6}$ and the $45_{+2}$ component plays a role in the
unified description of quark and charged lepton mass matrices $\cite{Matsuda}$.
Of course, there are no dangerous terms for the proton decay
in the coupling
$\overline{\psi}_{\overline{5}}^c \psi_{\overline{5}} 
\phi_{15}$.

However, the present model with a triplet scalar $\phi$ causes another 
problem, because the neutral component $\phi^0$ generally takes non-vanishing 
value or the vacuum expectation value $\langle \phi^0 \rangle$ . 


\section{SUMMARY}

The embedding of the $R$-parity violating SUSY models
into a GUT scenario, except for a special case, 
is ruled out because of the proton decay.
The SU(5) 10$+\overline{10}$ model \cite{Haba},
which is safe for the proton decay,
is ruled out if the solar neutrino
data establish $\sin^2 2\theta_{solar} \sim 0.8$.
On the other hand, an SO(10) 126 model is promising if neutrino
data show $\sin^2 2\theta_{solar} \sim 0.8$.
However, such a model with a triplet scalar
will cause another problems.
More study is required.
An SUSY GUT model with restricted $R$-parity violating 
which was discussed in 2.2 is also interesting.
The details will be given elsewhere.


\end{document}